\documentclass[twocolumn,showpacs,aps,prl,superscriptaddress,USletter]{revtex4}
\usepackage{graphicx}
\usepackage{dcolumn}
\usepackage{amsmath}
\usepackage{epsfig}

\input pubboard/babarsym.tex

\newcommand{\romegaKstz}{\ensuremath{2.4\pm1.1\pm0.7}}
\newcommand{\romegaKstp}{\ensuremath{0.6^{+1.4+1.1}_{-1.2-0.9}}}
\newcommand{\romegarhop}{\ensuremath{10.6\pm2.1^{+1.6}_{-1.0}}}
\newcommand{\romegarhoz}{\ensuremath{-0.6\pm0.7^{+0.8}_{-0.3}}}
\newcommand{\romegafz}{\ensuremath{0.9\pm0.4^{+0.2}_{-0.1}}}
\newcommand{\romegaomega}{\ensuremath{1.8^{+1.3}_{-0.9}\pm0.4}}
\newcommand{\romegaphi}{\ensuremath{0.1\pm0.5\pm0.1}}
\newcommand{\ulomegaKstz}{\ensuremath{4.2}}
\newcommand{\ulomegaKstp}{\ensuremath{3.4}}
\newcommand{\ulomegarhoz}{\ensuremath{1.5}}
\newcommand{\ulomegaomega}{\ensuremath{4.0}}
\newcommand{\ulomegaphi}{\ensuremath{1.2}}
\newcommand{\ulomegafz}{\ensuremath{1.5}}
\newcommand{\fLomegarhop}{\ensuremath{0.82\pm0.11\pm0.02}}
\newcommand{\Aomegarhop}{\ensuremath{0.04\pm 0.18\pm 0.02}}
\newcommand{\somegarhop}{\ensuremath{5.7}}

\newcommand{\BABARPubYear}    {06}
\newcommand{\BABARPubNumber}  {029}
\newcommand{\SLACPubNumber} {11848}

\newcommand{\calP}{\ensuremath{{\cal P}}}

\newcommand{\pvec}{{\bf p}}

\newcommand{\acp}{\ensuremath{\calA_{\CP}}}
\newcommand{\calB}{\ensuremath{{\cal B}}}

\newcommand{\timesix}{\ensuremath{\times10^{6}}}

\newcommand{\DE}{\ensuremath{\Delta E}}

\newcommand{\xf}{\ensuremath{{\cal F}}}

\newcommand{\thetaT}{\ensuremath{\theta_{\rm T}}}
\newcommand{\costhr}{\ensuremath{\cos\thetaT}}

\newcommand\etal{{\it et al.}}
\newcommand{\half}{\ensuremath{{1\over2}}}

\newcommand{\bfig}{\begin{figure}[htbpc!]}
\newcommand{\efig}{\end{figure}}
\newcommand\bef{\begin{figure}}
\newcommand\edf{\end{figure}}
\newcommand\dbline{\noalign{\vskip 0.10truecm\hrule}\noalign{\vskip 2pt}\noalign{\hrule\vskip 0.10truecm}}
\providecommand{\tbline}{\noalign{\vskip 0.05truecm\hrule\vskip0.05truecm}}

\newcommand\beq{\begin{equation}}
\newcommand\eeq{\end{equation}}
\newcommand\bear{\begin{array}}
\newcommand\enar{\end{array}}
\newcommand\beqa{\begin{eqnarray}}
\newcommand\eeqa{\end{eqnarray}}
\newcommand\ben{\begin{enumerate}}
\newcommand\een{\end{enumerate}}

\newcommand{\omtoppp}{\ensuremath{{\omega\ra\pip\pim\piz}}}

\newcommand{\Kst}{\ensuremath{K^*}}
\newcommand{\Kstp}{\ensuremath{\Kstarp}}
\newcommand{\Kstz}{\ensuremath{\Kstarz}}
   \newcommand{\KstpKppiz}{\ensuremath{\Kstarp_{K^+\pi^0}}}
   \newcommand{\KstptoKppiz}{\ensuremath{\Kstarp\ra K^+\pi^0}}
   \newcommand{\KstpKspip}{\ensuremath{\Kstarp_{\KS\pi^+}}}
   \newcommand{\KstptoKspip}{\ensuremath{\Kstarp\ra\KS\pi^+}}
   \newcommand{\KstzKppim}{\ensuremath{\Kstarz_{K^+\pi^-}}}
   \newcommand{\KstztoKppim}{\ensuremath{\Kstarz\ra K^+\pi^-}}

\newcommand{\kzs}{\ensuremath{\KS}}

\newcommand{\UfourS}{\ensuremath{\Upsilon(4S)}}

\providecommand{\fomegaKstp}{\ensuremath{\omega\Kstp}}
\providecommand{\omegaKstp}{\ensuremath{\Bp\ra\fomegaKstp}}
\newcommand{\BomegaKstp}{\ensuremath{\calB(\omegaKstp)}}

\providecommand{\fomegaKstz}{\ensuremath{\omega\Kstz}}
\providecommand{\omegaKstz}{\ensuremath{\Bz\ra\fomegaKstz}}
\newcommand{\BomegaKstz}{\ensuremath{\calB(\omegaKstz)}}

\newcommand{\fomegaKstpKspip}{\ensuremath{\omega\Kstp_{\KS \pip}}}

\newcommand{\fomegaKstpKppiz}{\ensuremath{\omega \Kstp_{\Kp\piz}}}

\newcommand{\fomegarhop}{\ensuremath{\omega\rho^+}\xspace}
\newcommand{\omegarhop}{\ensuremath{\Bp\ra\fomegarhop}\xspace}
\newcommand{\Bomegarhop}{\ensuremath{\calB(\omegarhop)}\xspace}

\newcommand{\fomegarhoz}{\ensuremath{\omega\rho^0}}
\newcommand{\omegarhoz}{\ensuremath{\Bz\ra\fomegarhoz}}
\newcommand{\Bomegarhoz}{\ensuremath{\calB(\omegarhoz)}}

\newcommand{\fomegaomega}{\ensuremath{\omega\omega}\xspace}
\newcommand{\omegaomega}{\ensuremath{\Bz\ra\fomegaomega}\xspace}
\newcommand{\Bomegaomega}{\ensuremath{\calB(\omegaomega)}\xspace}
\newcommand{\fomegaphi}{\ensuremath{\omega\phi}\xspace}
\newcommand{\omegaphi}{\ensuremath{\Bz\ra\fomegaphi}\xspace}
\newcommand{\Bomegaphi}{\ensuremath{\calB(\omegaphi)}\xspace}

\newcommand{\fomegafz}{\ensuremath{\omega f_0}\xspace}
\newcommand{\omegafz}{\ensuremath{\Bz\ra\fomegafz}\xspace}
\newcommand{\Bomegafz}{\ensuremath{\calB(\omegafz)}\xspace}

\def\sss{\scriptscriptstyle}
\def\barpd{{\raise.35ex\hbox{${\sss (}$}}--{\raise.35ex\hbox{${\sss )}$}}}
\def\BorBbar{\hbox{$B$\kern-0.85em\raise1.5ex\hbox{\barpd}\hspace{-0.4mm}$^0$}}

\def\figurebox#1#2#3{%
    \def\arg{#3}%
    \ifx\arg\empty
    {\hfill\vbox{\hsize#2\hrule\hbox to #2{\vrule\hfill\vbox to #1{\hsize#2\vfill}\vrule}\hrule}\hfill}%
    \else
    {\hfill\epsfbox{#3}\hfill}%
    \fi}

\begin{document}

\preprint{\babar-PUB-\BABARPubYear/\BABARPubNumber}
\preprint{SLAC-PUB-\SLACPubNumber}

\begin{flushleft}
\babar-PUB-\BABARPubYear/\BABARPubNumber\\
SLAC-PUB-\SLACPubNumber\\
\end{flushleft}

\title{
{\large \boldmath \bf  $B$ Meson Decays to $\omega\Kstar$, $\omega\rho$,  $\omega\omega$, $\omega\phi$, and $\omega f_0$ }
}

%
\author{B.~Aubert}
\author{R.~Barate}
\author{M.~Bona}
\author{D.~Boutigny}
\author{F.~Couderc}
\author{Y.~Karyotakis}
\author{J.~P.~Lees}
\author{V.~Poireau}
\author{V.~Tisserand}
\author{A.~Zghiche}
\affiliation{Laboratoire de Physique des Particules, F-74941 Annecy-le-Vieux, France }
% ...
\collaboration{The \babar\ Collaboration}
\noaffiliation

\date{\today}

\begin{abstract}
We describe searches for $B$ meson decays to the charmless vector-vector
final states $\omega\Kst$, $\omega\rho$, $\omega\omega$, and $\omega\phi$ 
with 233\timesix\ \BB\ pairs produced 
in \epem\ annihilation at $\sqrt{s}=10.58\ \gev$. We also search for the
 vector-scalar $B$ decay to $\omega f_0$.
 We measure the following branching fractions in units of $10^{-6}$:
 $\BomegaKstz = \romegaKstz ~~(<\ulomegaKstz)$, $\BomegaKstp =
 \romegaKstp ~~(<\ulomegaKstp)$, $\Bomegarhoz = \romegarhoz
 ~~(<\ulomegarhoz)$, $\Bomegarhop = \romegarhop$, $\Bomegaomega =
 \romegaomega ~~(<\ulomegaomega)$, $\Bomegaphi = \romegaphi
 ~~(<\ulomegaphi)$, and $\Bomegafz = \romegafz ~~(<\ulomegafz)$.  In each case the
 first error quoted is statistical, the second systematic, and the
 upper limits are defined at the 90\%\ confidence level.  For
 \omegarhop\ decays we also measure the longitudinal spin component 
 $f_L=\fLomegarhop$ and the charge asymmetry $\acp=\Aomegarhop$.
\end{abstract}

\pacs{13.25.Hw, 12.15.Hh, 11.30.Er}

\maketitle

Until recently, hadronic decays of $B$ mesons to pairs of light vector
mesons (VV final states) have received less theoretical and
experimental attention than decays to two pseudoscalar mesons (PP) or
one pseudoscalar and one vector meson (PV).  Early papers presented
calculations for branching fractions, $CP$-violating asymmetries
\cite{VVBRAcp}, and relative spin component contributions~\cite{VVPol}
for these decays.  The measurement three years ago of an unexpectedly
small value of the fraction of the longitudinal spin component ($f_L$)
in penguin-dominated $B\to\phi\Kstar$
decays~\cite{BabarPhiKstar,BellePhiKstar} triggered new theoretical
activity.  There have been several attempts to understand the small
value of $f_L$ within the Standard Model (SM)~\cite{VVBSMrefs} and
many papers suggested non-SM explanations~\cite{nSMetc}.  Further
information about these effects can come from both branching fraction
and $f_L$ measurements in decays such as $B\to\omega\Kstar$ or
$B\to\omega\phi$, which are conjugate to $B \to \phi\Kstar$ via an
SU(3) rotation~\cite{oh}.  Information on these and related charmless
decays can additionally be used to provide sensitivity to the CKM
angles $\alpha$ and $\gamma$~\cite{VVphaserefs}.

We have discussed above mostly penguin-dominated decays.  There
are also decays with the $\Kstar$ replaced by a $\rho$, $\omega$ or 
$\phi$ meson where tree diagrams are expected to be more important.
These include $B$ decays to the final states $\rho\rho$, $\omega\rho$,
and $\omega\omega$.  The decay $B\to\rho\rho$ is known to
be nearly fully longitudinally polarized~\cite{BaBarVV03,BelleRhoRho0Obs} 
and these other predominantly tree decays are expected to behave similarly~\cite{lilu},
with branching fractions as predicted in~\cite{VVBRAcp}.

We report results of measurements of $B$ decays to the charmless VV
final states $\omega V$, where $V$ represents a neutral or charged
\Kstar or $\rho$, or an $\omega$ or $\phi$ meson. We also measure the
decay $\Bz\to\omega f_0(980)$ which shares the same final state as the
$\Bz\to\omega\rho^0$ decay.  Due to the current small signal samples,
only the branching fractions and the fraction of the longitudinal spin 
component are measured, the latter by
integrating over the azimuthal angles, for which the azimuthal
acceptance is uniform.  The angular distribution is
\begin{eqnarray}
    &&\frac{1}{\Gamma}\frac{d^2\Gamma}{d\cos{\theta_1}d\cos{\theta_2}}
    =     \label{eq:vvAngDist} \\
&&\frac{9}{4}\left\{\frac{1}{4}(1-f_L)\sin^2{\theta_1}\sin^2{\theta_2}
     +f_L\cos^2{\theta_1}\cos^2{\theta_2}\right\}\ , \nonumber
\end{eqnarray}
where $\theta_k$ is the helicity angle in the $V_k$
rest frame with respect to the boost axis  from the $B$ 
rest frame and $f_L$ is the fraction of the longitudinal spin component.
For \omegarhop, we
also measure the direct \CP-violating time-integrated charge asymmetry
$\acp = (\Gamma^- - \Gamma^+)/(\Gamma^- + \Gamma^+)$,
where the superscript on the $\Gamma$ corresponds to the sign of the \Bpm meson.


The results presented here are based on data collected
with the \babar\ detector~\cite{BABARNIM}
at the PEP-II asymmetric $e^+e^-$ collider
located at the Stanford Linear Accelerator Center.  An integrated
luminosity of 211~fb$^{-1}$, corresponding to 
232.8\timesix\ \BB\ pairs, was recorded at the $\Upsilon (4S)$
resonance (center-of-mass energy $\sqrt{s}=10.58\ \gev$).

Charged particles from the \epem\ interactions are detected, and their
momenta measured, by five layers of double-sided
silicon microstrip detectors surrounded by a 
40-layer drift chamber, both operating in the 1.5-T magnetic
field of a superconducting solenoid. We identify photons and electrons 
using a CsI(Tl) electromagnetic calorimeter (EMC).
Further charged particle identification (PID) is provided by the average energy
loss ($dE/dx$) in the tracking devices and by an internally reflecting
ring-imaging Cherenkov detector (DIRC) covering the central region.

We reconstruct the $B$-daughter candidates through their decays
$\rho^0\ra\pip\pim$, $f_0(980)\ra\pip\pim$, $\rho^+\ra\pip\piz$,
\KstztoKppim (denoted \KstzKppim), \KstptoKppiz (\KstpKppiz), \KstptoKspip
(\KstpKspip), \omtoppp, $\phi \ra K^+K^-$, $\piz\ra\gaga$, and
$\kzs\ra\pip\pim$ (charge-conjugate decay modes are implied
throughout).  Table \ref{tab:rescuts}\ lists the requirements on the
invariant mass of these particles' final states.  For the $\rho$,
\Kstar, $\phi$, and $\omega$ invariant masses these requirements are
set loose enough to include sidebands, as these mass values are
treated as observables in the maximum-likelihood fit described below.
The $\rho^0$ and $f_0$ (we use $f_0$ as shorthand for $f_0(980)$)
yields are extracted from a simultaneous fit to the same data sample.
For \kzs\ candidates we further require the three-dimensional flight
distance from the event primary vertex to be greater than three times
its uncertainty.  Secondary pion and kaon candidates in $\rho$,
\Kstar, and $\omega$ candidates are rejected if their DIRC, $dE/dx$,
and EMC PID signature satisfies tight consistency with protons or
electrons, and the kaons (pions) must (must not) have a kaon
signature.

\begin{table}[btp]
\begin{center}
\caption{
Selection requirements on the invariant mass and helicity angle
of $B$-daughter resonances.  The helicity angle is unrestricted unless indicated otherwise.
}
\label{tab:rescuts}
\begin{tabular}{lcc}
\dbline
State           & inv. mass (\mev)  &    helicity angle                     \\
\dbline                                        
\KstzKppim,\KstpKspip&$755 < m_{K\pi}< 1035$   &$-0.85<\cos{\theta}<1.0$\\
\KstpKppiz      &$755 < m_{K\pi}< 1035$  &$-0.8<\cos{\theta}<1.0$        \\
$\rho^0/f_0$    & $540 < m_{\pi\pi} <1060$       &$-0.85<\cos{\theta}<0.85$\\
$\rho^+$        & $470 < m_{\pi\pi} <1070$       &$-0.7<\cos{\theta}<0.85$\\
$\omega$        & $740 < m_{\pi\pi\pi} <820$     &    \\
$\phi$            & $1009<m_{KK} <1029$          &       \\
\piz            & $120 < m_{\gamma\gamma} < 150$ &       \\
\kzs            & $473<m_{\pi\pi} <522$          &       \\
\dbline
\end{tabular}
\vspace{-5mm}
\end{center}
\end{table}

Table \ref{tab:rescuts}\ also gives the restrictions on the \Kstar and
$\rho$ helicity angles $\theta$ (previously defined for Eq.~\ref{eq:vvAngDist}),
imposed to avoid regions of rapid acceptance variation or
combinatorial background from soft particles.  
To calculate $\theta$ we take the angle relative to,  for $\omega$, 
the helicity axis of the normal to the decay plane, for $\rho$ and
$\phi$, the positively-charged (or only charged) daughter momentum,
and for \Kstar\ the daughter kaon momentum.

A $B$-meson candidate is characterized kinematically by the
energy-substituted mass $\mes=\sqrt{(\half
s+\pvec_0\cdot\pvec_B)^2/E_0^{*2}-\pvec_B^2}$ and the energy
difference $\DE = E_B^*-\half\sqrt{s}$, 
where $(E_0,\pvec_0)$ and $(E_B,\pvec_B)$ are four-momenta of
the \UfourS\ and the $B$ candidate, respectively, $s$ is the square of
the center of mass energy,
and the asterisk denotes the \UfourS\ frame. The resolution on \DE\
(\mes) is about 30 MeV ($3.0\ \mev$).
Our signal falls in the region $|\DE|\le0.1$ GeV and
$5.27\le\mes\le5.29\ \gev$, which is then extended to include sidebands
that provide a good description of the background.  The
average number of candidates found per selected event in data is in
the range 1.1 to 1.7, depending on the final state.  We choose the
candidate with the smallest value of a $\chi^2$ constructed from the
deviations of the $B$ daughter resonance masses (all particles in
Table \ref{tab:rescuts} except \piz, \KS, and $f_0$) 
from their expected~\cite{PDG2004} values.

Backgrounds arise primarily from random combinations of particles in
continuum $\epem\ra\qqbar$ events ($q=u,d,s,c$).  We reduce these by
using the angle \thetaT\ between the thrust axis of the $B$
candidate in the \UfourS\ frame and that of the rest of the charged
and neutral particles in the event.  The
distribution of $|\costhr|$ is sharply peaked near $1.0$ for
\qqbar\ jet pairs, and nearly uniform for
$B$ meson decays.  The requirements, chosen to reduce the sample size
for the large background modes, are $|\costhr|<0.9$ for the $\omega\phi$ mode,
$|\costhr|<0.8$ for the $\omega\omega$ and  $\omega\Kstar$ modes, and $|\costhr|<0.7$ for the
$\omega\rho$ modes.  In the maximum-likelihood fit described below, 
we also use a Fisher
discriminant \xf\ that combines four variables defined
in the \UfourS\ frame: the polar angles with respect to the beam axis of the
$B$ momentum and $B$ thrust axis, and the zeroth and second angular
moments, $L_{0}$ and $L_{2}$, of the energy flow about the $B$ thrust axis in the \UfourS\ frame.  The moments are defined by $ L_j = \sum_i
p_i\times\left|\cos\theta_i\right|^j,$ where $\theta_i$ is the angle
with respect to the $B$ thrust axis of a charged or neutral particle $i$,
$p_i$ is its momentum, and the sum excludes the $B$ candidate's
daughters.

From Monte Carlo (MC) simulation \cite{geant} we determine the most
important charmless \BB\ backgrounds (typically about a dozen background
modes for each signal
final state).  We include a variable yield for these in the fit
described below.  We also introduce a component for non-resonant
$\pi\pi$ and $K\pi$ background.  The magnitude of this component is
fixed in the fit as determined from extrapolations from higher-mass
regions.

We obtain yields and values of $f_L$ and \acp from extended unbinned 
maximum-likelihood fits with input observables \DE, \mes, \xf and, for the
vector meson $k$, the mass $m_k$ and helicity  angle $\theta_k$.
For each event $i$ and hypothesis $j$ (signal, \qqbar\ background, 
\BB\ background) we define the probability density function (PDF)
\begin{equation}
\calP^i_j = \calP_j(\mes^i) \calP_j(\DE^i) 
 \calP_j(\xf^i) \calP_j(m_1^i,m_2^i,\theta^i_1,\theta^i_2).
    \label{eq:evtL}
\end{equation}
We check for correlations in the background observables beyond those
accounted for in this PDF and find them to be small.  For the signal
component, we correct for the effect of small neglected correlations 
(see below).  The likelihood function is
\begin{equation}
    {\cal L} = \frac{e^{-(\sum Y_j)}}{N!} \prod_{i=1}^N
\sum_j Y_j {\cal P}_j^i\ , 
    \label{eq:totalL}
\end{equation}
where $Y_j$ is the yield of events of hypothesis $j$
and $N$ is the number of events in the sample.  
  
\begin{table*}[!bth]
\caption{
Signal yield $Y$ and its statistical uncertainty, bias $Y_0$,
detection efficiency $\epsilon$, daughter branching fraction product
$\prod\calB_i$, significance $S$ (with systematic uncertainties included), 
measured branching fraction \calB, 90\% C.L. upper limit, measured
or assumed longitudinal polarization, and \acp.
}
\label{tab:results}
\begin{tabular}{lccrcccccc}
\dbline
Mode            & $Y$      &$Y_0$       &$\epsilon$~~ &$\prod\calB_i$ & $S$       &  \calB        & \calB\ U.L. &  ~~$f_L$  & \acp \\
                & (events) & (events)   &(\%)~& (\%) &~~($\sigma$)~~ & $(10^{-6})$   & $(10^{-6})$ &   & \\
\dbline
\fomegaKstz    & $55^{+20}_{-19}$ &$11$    &13.2 &59.2 &2.4 & \romegaKstz & \ulomegaKstz & $0.71^{+0.27}_{-0.24}$ & --- \\
\tbline
\fomegaKstpKspip&  $-3.6^{+10}_{-8}$&$-5$    &12.5 &20.3 &0.1 & ~${0.2^{+1.7}_{-1.5} ~^{+1.5}_{-1.1}}$   & 4.2 & 0.7 fixed & --- \\
\fomegaKstpKppiz& $12^{+14}_{-12}$&$6$     & 8.0 &29.6 &0.5 & ${1.1^{+2.5+1.3}_{-2.1-1.2}} $  & 5.7 & 0.7 fixed  & --- \\
\fomegaKstp    &                  &            &     &     &0.4 & \romegaKstp & \ulomegaKstp &0.7 fixed     & --- \\
\tbline
\fomegarhoz    & $-18^{+17}_{-14}$&$-2$    &11.6 &89.1 &0.6 & \romegarhoz & \ulomegarhoz &0.9 fixed      & --- \\
\tbline
\fomegafz      & $25^{+12}_{-11}$ &$4$     &15.2 &59.4 &2.8 & \romegafz   & \ulomegafz& ---       & --- \\
\tbline
\fomegarhop    & $156^{+33}_{-31}$&$11$    & 6.6 &89.1 &\somegarhop & \romegarhop & --- & \fLomegarhop  & ~~\Aomegarhop \\
\tbline
\fomegaomega   & $48^{+24}_{-19}$ &$8$     &12.9 &77.5 &2.1 & \romegaomega& \ulomegaomega &0.79$\pm0.34$ & --- \\
\tbline
\fomegaphi     &$3.1^{+4.4}_{-8.5}$&$1$&19.0 &43.2 &0.3 & \romegaphi  & \ulomegaphi & 0.88 fixed & --- \\
\dbline
\end{tabular}
\vspace{-5mm}
\end{table*}

The PDF factor for the resonances in the signal takes the form
$\calP_{1,{\rm sig}}(m_1^i)\calP_{2,{\rm sig}}(m_2^i){\cal
Q}(\theta^i_1,\theta^i_2)$ with ${\cal Q}$ given by Eq.\
\ref{eq:vvAngDist}\ modified to account for detector acceptance.  For 
\qqbar\ background it is given for each resonance independently by
$\calP_{\qqbar}(m_k^i,\theta_k^i) = \calP_{\rm pk}(m_k^i)\calP_{\rm
pk}(\theta_k^i) + \calP_{\rm c}(m_k^i)\calP_{\rm c}(\theta_k^i)$,
distinguishing between true resonance ($\calP_{\rm pk}$) and
combinatorial ($\calP_{\rm c}$) components.  For 
the \BB\ background we assume that all four mass and helicity angle observables
are independent.  

For the signal, \BB\ background, and non-resonant background 
components we determine the PDF
parameters from simulation.
We study large control samples of $B\to D\pi$ decays of similar
topology to verify the simulated resolutions in \DE\ and \mes,
adjusting the PDFs to account for any differences found.
  For the continuum background we use
(\mes,\,\DE) sideband data to obtain initial values, before applying
the fit to data in the signal region, and leave them free
to vary in the final fit.

The parameters that are allowed to vary in the fit include the signal,
\BB\ background, and non-resonant background yields, and continuum
background PDF parameters. For the three modes with signals of more
than $2\sigma$ significance, we vary $f_L$ in the fit 
to properly account for the
variation of efficiency with $f_L$.  For \omegarhop\ we also vary the
signal and background charge asymmetries.  For the fits with little
signal, we fix $f_L$ to a value that is consistent with \textit{a
  priori} expectations (see Table \ref{tab:results}), and account for
the associated systematic uncertainty.  

To describe the PDFs, we use the sum of two Gaussians for ${\cal
  P}_{\rm sig}(\mes)$, ${\cal P}_{\rm sig}(\DE)$, and the true
resonance components of ${\cal P}_j(m_k)$; for \xf\ we use an
asymmetric Gaussian for signal with a small Gaussian component for
background to account for important tails in the signal region.  The
background \mes\ shape is described by the function
$x\sqrt{1-x^2}\exp{\left[-\xi(1-x^2)\right]}$ (with
$x\equiv\mes/E_B^*$) and the distributions of masses $m_k$ by second
or third order polynomials.  The background PDF parameters that are
allowed to vary in the fit are $\xi$ for \mes, slope for \DE, the
polynomial describing the combinatorial component for $m_k$, and the
peak position and lower and upper width parameters for \xf.

We evaluate possible biases from our neglect of correlations among
discriminating variables in the PDFs by fitting ensembles of simulated
experiments into which we have embedded the expected number of signal
events and \BB\ background events, randomly extracted 
from the fully-simulated MC samples.  We give
in Table \ref{tab:results} the yield bias $Y_0$ found for each
mode.  Since events from a weighted mixture of simulated \BB\ background
decays are included, the bias we measure includes the effect of
this background.

The systematic uncertainties on the branching fractions arising from
lack of knowledge of the PDFs have been included in part in the
statistical error since most background parameters are free in the
fit.  For the signal, the uncertainties in PDF parameters are
estimated from the consistency of fits to MC and data in large control
samples of topology similar to signal.  Varying the signal PDF
parameters within these errors, we estimate yield uncertainties for
each mode.  The uncertainty in the yield bias correction is taken to
be the quadratic sum of two terms: half the bias correction and the
statistical uncertainty on the bias itself.  Similarly, we estimate
the uncertainty from modeling of the \BB\ backgrounds as the
change in the signal yield when 
the number of fitted \BB\ events is
fixed to be within one sigma of the expected number of \BB\ events
from MC.  For the non-resonant $\pi\pi$ or $K\pi$ backgrounds the
uncertainty is taken as the change in the signal when the background
yield is varied within the uncertainty of the fits to the higher-mass
regions.  For modes with fixed $f_L$, the uncertainty due to the
dependence of signal efficiency on $f_L$ is evaluated as the measured
change in branching fraction when $f_L$ is varied by $\pm 0.3$ (up to
a maximum of $f_L=1$).  These additive systematic errors are dominant
for all modes; the PDF variation is always the smallest but the others
are typically similar in size.

Uncertainties in our knowledge of the efficiency, found from studies
of data control samples, include $0.8\%\times N_t$, $3.0\%\times
N_{\pi^0}$, and 1\%\ for a \KS\ decay, where $N_t$ is the number of
tracks, and $ N_{\pi^0}$ is the number of $\pi^0$s in a decay.
  We estimate the
uncertainty in the number of $B$ mesons to be 1.1\%.  Published data
\cite{PDG2004}\ provide the uncertainties in the $B$-daughter product
branching fractions (1--2\%).  The uncertainties from the event
selection are 1--2\%\ for the requirement on \costhr.

The systematic uncertainty on $f_L$ for \omegarhop\ includes the
effects of fit bias, PDF-parameter variation, and \BB\ and non-resonant
backgrounds, all estimated with the same method as for the yield
uncertainties described above.  From large inclusive kaon and
$B$-decay samples, we find a systematic uncertainty for \acp\ of
$0.02$ due mainly to the dependence of reconstruction efficiency on the 
momentum of the charged $\rho$ daughter. We find for the \omegarhop\
background, $\acp^{qq}=-0.010\pm0.007$, confirming this estimate.

In Table \ref{tab:results} we also show for each decay mode the
measured branching fraction with its uncertainty and significance
together with the quantities entering into these computations.  The
significance is taken as the square root of the difference between the
value of $-2\ln{\cal L}$ (with systematic uncertainties included) for
zero signal and the value at its minimum.  For all modes except for
\omegarhop\ we quote a 90\%\ confidence level (C.L.) upper limit,
taken to be the branching fraction below which lies 90\% of the total
of the likelihood integral in the positive branching fraction region.
In calculating branching fractions we assume that the decay rates of
the \UfourS\ to \BpBm\ and \BzBzb\ are equal \cite{PDG2004}.  For
decays with \Kstarp, we combine the results from the two \Kstar\ decay
channels, by adding their values of $-2\ln{\cal L}$, taking into
account the correlated and uncorrelated systematic errors.

\begin{figure}[!htbp]
\begin{center}
  \includegraphics[width=1.0\linewidth]{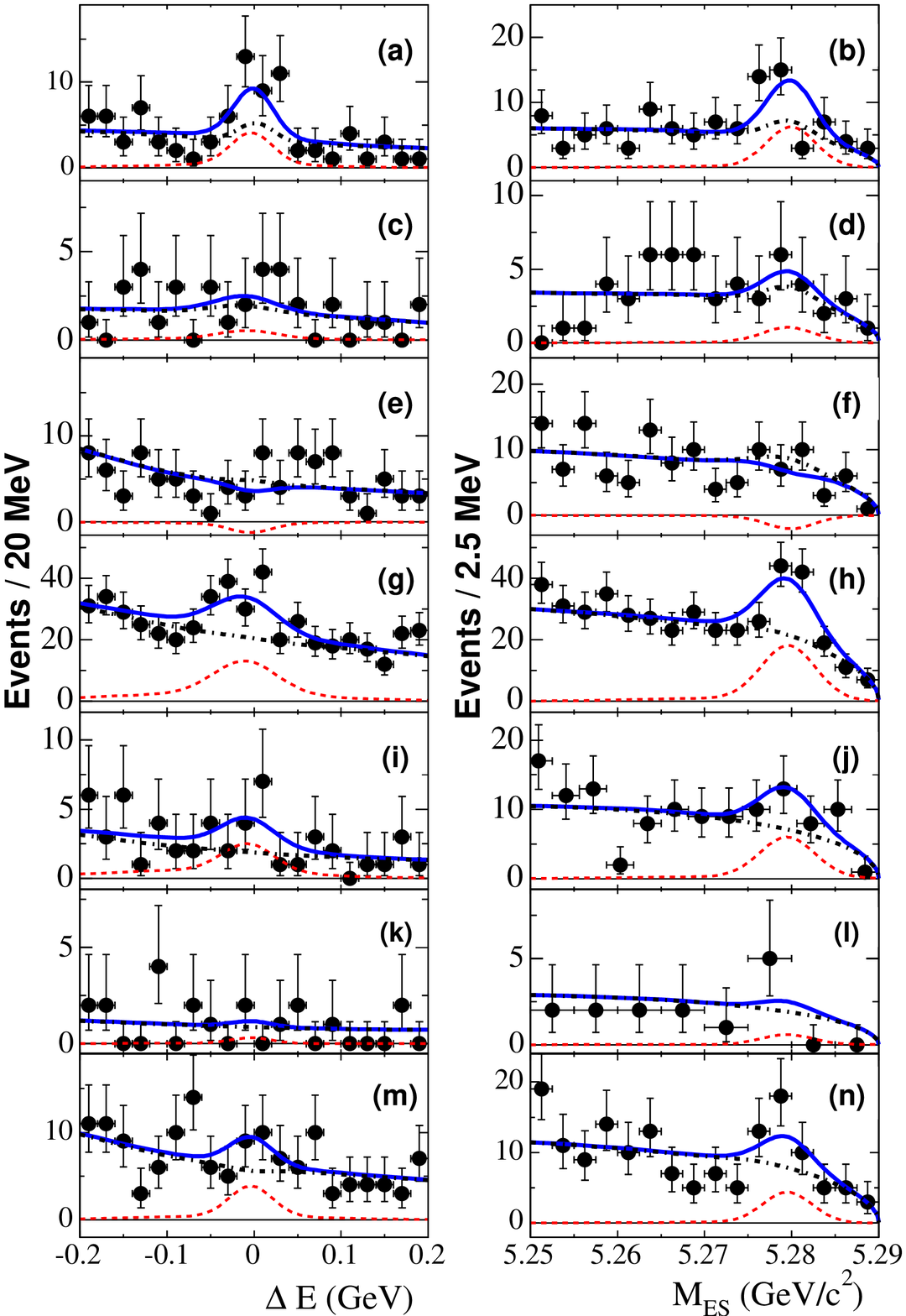}
  \caption{Projections of  \DE\ (left) and \mes\ 
    (right) of events passing a signal likelihood 
    threshold for \omegaKstz, (a,b), \omegaKstp, (c,d) \omegarhoz,
    (e,f) \omegarhop, (g,h) \omegaomega, (i,j) \omegaphi, (k,l) and
    \omegafz, (m,n).  The solid curve is the fit function, 
the dashed curve is the signal
    contribution, and the dot-dashed curve is the background
    contribution.  }
  \label{fig:proj_omegaKst}
\end{center}
\end{figure}

\begin{figure}[!htbp]
\begin{center}
  \includegraphics[width=1.0\linewidth]{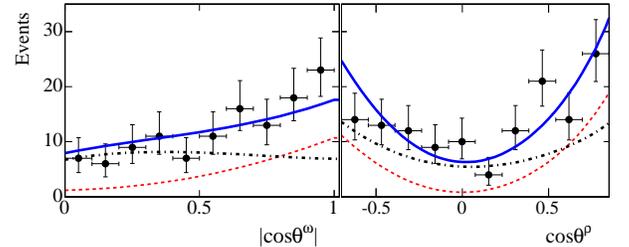}
  \caption{Projections of helicity-angle cosines, 
    of events passing a signal likelihood  threshold for
    $\omega$ (left) and $\rho^+$ (right) from the fit for \omegarhop
    decay.  The solid curve is the fit function, 
the dashed curve is the signal contribution,
    and the dot-dashed curve is the background contribution.  }
  \label{fig:sPlot_hel}
\end{center}
\end{figure}

We present in
Fig.\ \ref{fig:proj_omegaKst}\  the data and PDFs projected
onto \mes\ and \DE, for subsamples enriched with a mode-dependent
threshold requirement on the  signal likelihood
(computed without the PDF associated with the variable plotted) chosen
to optimize the significance of signal in the resulting subsample.
Fig.\ \ref{fig:sPlot_hel}\ gives projections
of $\cos\theta$ for the \omegarhop\ decay.

\begin{figure}[htbp]
\hspace{-0.3cm}
 \scalebox{1.0}{
  \includegraphics[width=1.0\linewidth, angle=-0]{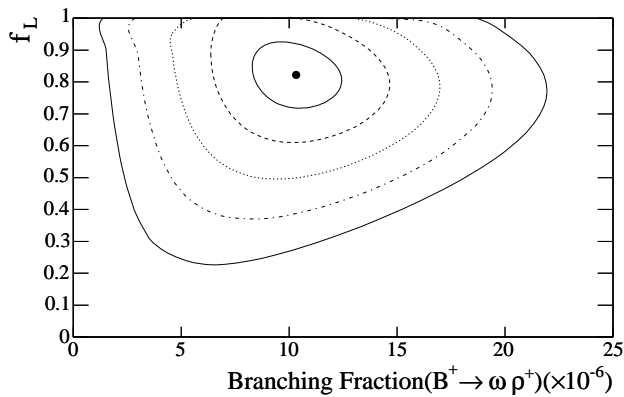}
 }
  \caption{Distribution of $-2\ln{\calL}(f_L,\calB)$ for \omegarhop
 decay.
The solid dot gives the central value; curves give the contours in
1-sigma steps ($\Delta\sqrt{-2\ln{\calL}(f_L,\calB)}=1$).}  
  \label{fig:contour_omegaRho}
\end{figure}

The branching fraction value \calB\ given in Table \ref{tab:results}\
for \omegarhop\ comes from a direct fit with the free parameters \calB\
and $f_L$, as well as \acp.  This choice exploits the feature that
\calB\ is less correlated with $f_L$ than is either the yield or
efficiency taken separately.  The behavior of $-2\ln{\calL}(f_L,\calB)$
is shown in Fig.~\ref{fig:contour_omegaRho}.

In summary, we have searched for seven charmless hadronic $B$-meson decays.
We observe \omegarhop\ with a significance of
\somegarhop\ $\sigma$, and establish improved 90\% C.L. upper limits
for the other modes, with the following branching fractions:
\begin{eqnarray}
    \BomegaKstz & = & \romegaKstz ~~(<\ulomegaKstz)\times10^{-6} ,\nonumber\\
    \BomegaKstp & = & \romegaKstp ~~(<\ulomegaKstp)\times10^{-6},\nonumber\\
    \Bomegarhoz & = & \romegarhoz ~~(<\ulomegarhoz)\times10^{-6},\nonumber\\
    \Bomegarhop & = & \romegarhop \times10^{-6},\nonumber\\
    \Bomegaomega& = & \romegaomega ~~(<\ulomegaomega)\times10^{-6},\nonumber\\
    \Bomegaphi  & = & \romegaphi ~~(<\ulomegaphi)\times10^{-6},\; \mbox{and}\nonumber \\
    \Bomegafz   & = & \romegafz ~~(<\ulomegafz)\times10^{-6}.\nonumber
\end{eqnarray}
In each case the first error quoted is statistical, the second systematic, and
the upper limits are taken at 90\% C.L.

For \omegarhop\ we also measure the longitudinal spin component
$f_L=\fLomegarhop$ and charge asymmetry $\acp=\Aomegarhop$. The
longitudinal spin component is dominant, as it is for $B\to\rho\rho$
\cite{BaBarVV03,BelleRhoRho0Obs}. Assuming tree dominance we would
naively expect the branching fraction for \omegarhop\ to be equal to
that of $\Bp\to\rho^+\rho^0$. 
However the
measured branching fraction for \Bomegarhop\ is more than two
standard deviations smaller than the world average 
${\cal B}(\Bp\to\rho^+\rho^0)$=$26\pm 6\times10^{-6}$
~\cite{PDG2004}.

Our branching fraction results are in agreement with
theoretical estimates \cite{VVBRAcp,lilu}.

We are grateful for the excellent luminosity and machine conditions
provided by our \pep2\ colleagues, 
and for the substantial dedicated effort from
the computing organizations that support \babar.
The collaborating institutions wish to thank 
SLAC for its support and kind hospitality. 
This work is supported by
DOE
and NSF (USA),
NSERC (Canada),
IHEP (China),
CEA and
CNRS-IN2P3
(France),
BMBF and DFG
(Germany),
INFN (Italy),
FOM (The Netherlands),
NFR (Norway),
MIST (Russia), and
PPARC (United Kingdom). 
Individuals have received support from CONACyT (Mexico), 
Marie Curie EIF (European Union),
the A.~P.~Sloan Foundation, 
the Research Corporation,
and the Alexander von Humboldt Foundation.

\end{document}